\documentclass[prl,aps,superscriptaddress,twocolumn]{revtex4-1}
\usepackage{graphicx}
\usepackage{bm}
\usepackage{amssymb,amsfonts,textcomp}
\usepackage{multirow}
\usepackage{color}
\usepackage{array}
\usepackage{hhline}
\usepackage[normalem]{ulem}

\usepackage{amsmath}    
\usepackage{verbatim}   
\usepackage{subfigure}  

\begin{document}

\title{Superfluid $^4$He dynamics beyond quasiparticle excitations} 
\author{K. Beauvois} 
   \affiliation{Institut Laue-Langevin, 6, rue Jules Horowitz, 38042 Grenoble, France}
   \affiliation{Univ.\ Grenoble Alpes, Inst NEEL, F-38000 Grenoble, France}
   \affiliation{CNRS, Inst NEEL, F-38042 Grenoble, France}
\author{C. E. Campbell}
   \affiliation{School of Physics and Astronomy, University of Minnesota, Minneapolis MN 55455, USA}
\author{J. Dawidowski}
   \affiliation{Comisi\'on Nacional de Energ\'ia At\'omica and CONICET, Centro At\'omico Bariloche, (8400) San Carlos de Bariloche,
R\'io Negro, Argentina}
\author{B. F\aa k} 
   \affiliation{Institut Laue-Langevin, 6, rue Jules Horowitz, 38042 Grenoble, France}
   \affiliation{Univ.\ Grenoble Alpes, INAC-SPSMS, F-38000 Grenoble, France}
   \affiliation{CEA, INAC-SPSMS, F-38000 Grenoble, France} 
\author{H. Godfrin}
	 \email[]{henri.godfrin@neel.cnrs.fr}
   \affiliation{Univ.\ Grenoble Alpes, Inst NEEL, F-38000 Grenoble, France}
   \affiliation{CNRS, Inst NEEL, F-38042 Grenoble, France}
\author{E.~Krotscheck}
   \affiliation{Department of Physics, University at Buffalo, SUNY Buffalo NY 14260, USA}
   \affiliation{Institute for Theoretical Physics, Johannes Kepler University, A 4040 Linz, Austria}
\author{H.-J. Lauter}
   \affiliation{Spallation Neutron Source, Oak Ridge National Laboratory, Oak Ridge, Tennessee 37831, USA}
\author{T. Lichtenegger}
   \affiliation{Department of Physics, University at Buffalo, SUNY Buffalo NY 14260, USA}
   \affiliation{Institute for Theoretical Physics, Johannes Kepler University, A 4040 Linz, Austria}
\author{J. Ollivier}
   \affiliation{Institut Laue-Langevin, 6, rue Jules Horowitz, 38042 Grenoble, France}
\author{A. Sultan}
   \affiliation{Univ.\ Grenoble Alpes, Inst NEEL, F-38000 Grenoble, France}
   \affiliation{CNRS, Inst NEEL, F-38042 Grenoble, France}
   \affiliation{CEA, INAC-SPSMS, F-38000 Grenoble, France} 

\date{\today}

\begin{abstract}
 	The dynamics of superfluid $^4$He at and above the Landau
  quasiparticle regime is investigated by high precision inelastic
  neutron scattering measurements of the dynamic structure
  factor.  
	A highly structured response is observed above the
  familiar phonon-maxon-roton spectrum,   
  characterized by sharp thresholds for  
	phonon-phonon, maxon-roton and roton-roton coupling processes. 
	The experimental dynamic structure factor is compared to the  
  calculation of the {\em same\/} physical quantity 
	by a Dynamic Many-body theory	
	including three-phonon processes self-consistently. 
	The theory is found to provide a quantitative description 
	of the dynamics of the correlated bosons for energies up to about
	three times that of the Landau quasiparticles.
	
\end{abstract}

\pacs{05.30.Jp,67.25.dt,67.25.D-,78.70.Nx}

\maketitle 

Liquid $^4$He is the prime example of a strongly correlated quantum
many-body system. It has been studied for decades and still offers
surprises that lead to new insights. Understanding the helium fluids,
due to their generic nature, lies at the core of understanding other
strongly correlated many-particle systems, and is therefore of
interest not only for the quantum fluids community. The description of
the elementary excitations of superfluid $^4$He in terms of phonon-roton
quasiparticles is a cornerstone of modern physics, with profound
implications in condensed matter physics, particle physics and cosmology. 
The empirical dispersion relation of these excitations proposed by Landau
\cite{Landauroton,Landauroton2} to explain thermodynamic data found
support in the microscopic theory of Feynman and Cohen
\cite{FeynmanBackflow}, initiating a fruitful development of the field
theoretical description of correlated quantum particles. 

From an experimental point of view, neutron scattering
techniques allowed the direct observation of very sharp excitations in
superfluid $^4$He at low temperatures. The density fluctuations
displayed, as predicted, a continuous phonon-maxon-roton dispersion
curve: a linear phonon part at low wave vectors followed by a maximum
(``maxon'') and then a pronounced ``roton'' minimum at a finite
wave vector of atomic dimensions.  Phonons naturally arise as the
Goldstone mode associated with the continuous symmetry of the
interacting system, whereas rotons are a direct consequence of strong
correlations. Roton-like excitations have been proposed in cold
atomic gases \cite{PhysRevLett.109.235307}, in 
one-dimensional $^4$He \cite{1D-4He} and in two-dimensional
fermionic systems \cite{Nature2p2h}. Superfluidity emerges
phenomenologically as a natural consequence of the dynamics, while the
knowledge of the dispersion relation allows the calculation of low temperature 
thermodynamic properties of superfluid $^4$He
\cite{DonnellyDonnellyHills}.
 
The relation between theory and experiment, however, is not
straightforward \cite{PinesNoz,GlydeBook}.  The excitations considered
by Landau, Feynman and others correspond to the single-particle
response function associated with the description of an effective
vacuum - the superfluid ground state - and non-interacting
quasiparticle excitations.  Neutron scattering, in turn, gives access
to the dynamic structure factor $S(Q,\omega)$, a quantity related to the
dynamic susceptibility, {\em i.e.},  the linear response of the
system to a density fluctuation.  
The latter has a strong weight along the phonon-roton
dispersion curve, but it also contains additional 
contributions, already observed in early neutron scattering
experiments \cite{CowleyWoods,Andersen92,Andersen94a,AndersenRoton}. 
Much of the work on
superfluid $^4$He has been focused on the single-particle response
function, attempting to extract 
the quasiparticle dispersion relation \cite{Andersen94a} 
and life-time \cite{Mezei80,Fak12}
from neutron data.
In parallel, considerable effort has been devoted to the development of  
an accurate theoretical description  
of the dynamics of superfluid $^4$He using various  
techniques \cite{GlydeBook,TriesteBook,Moroni,Vitali,Arrigoni,eomIII}.

In this Letter, we investigate in detail the multi-excitation region 
of the dynamic structure factor $S(Q,\omega)$ of superfluid $^4$He. 
We present new, high-precision neutron scattering measurements of 
$S(Q,\omega)$ at very low temperatures. 
We find several new features, in particular a ``ghost phonon'',
but also multi-particle thresholds that are much sharper than in earlier 
experimental work. 
These features are in agreement with a quantitative  
microscopic calculation of the density fluctuations $S(Q,\omega)$ 
within  a recent Dynamic Many-Body theory \cite{eomIII}.
Theoretical and experimental results for $S(Q,\omega)$ in a broad sector 
of the spectrum can be compared directly, leading to an unprecedentedly 
accurate description of the dynamics of superfluid $^4$He.

The inelastic neutron scattering measurements were performed on the
neutron time-of-flight spectrometer IN5 at the Institut Laue-Langevin
using an incoming energy of 3.55\,meV (wavelength 4.8\,\AA)
and an energy resolution at
elastic energy transfer of 0.07\,meV.  The high-purity superfluid
$^4$He sample was contained in a thin-walled cylindrical aluminum
container of inner diameter 15\,mm. The effective sample height 
in the beam was 50\,mm.
Cadmium disks were placed inside 
the cell at intervals of 10\,mm to reduce multiple scattering, an 
important experimental artifact discussed below. 
The cell was connected to the mixing chamber of a dilution refrigerator 
via a copper piece equipped with silver sinter to ensure good thermal contact, 
thereby allowing measurements to be done at very low temperatures, $T<100$\,mK.
The measurements were performed at saturated vapor pressure. 

\begin{figure}
\includegraphics[width=1.0\columnwidth]{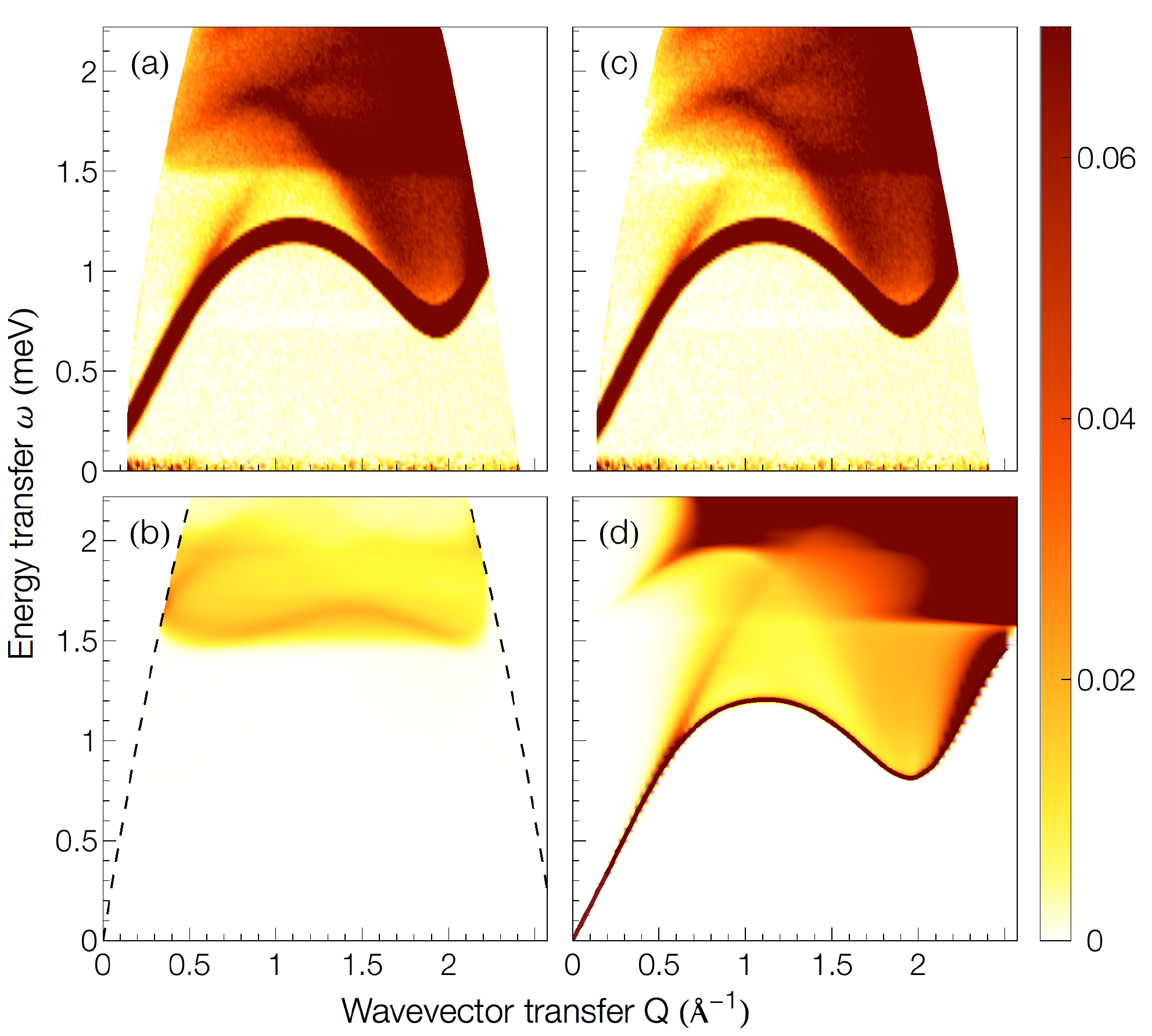}
\caption{(color online) (a)~$S(Q,\omega)$ of superfluid $^4$He
measured as a function of wave vector and energy transfer, 
at saturated vapor pressure and temperature $T\,\le\,100$\,mK. 
Contributions involving scattering with the aluminium cell 
have been subtracted, but not multiple scattering within the helium.
(b)~Helium multiple scattering contribution (numerical simulation); 
note that its magnitude is comparable to the multi-particle intensity 
seen in panels (c) and (d), and in Fig.\ \ref{Figure-3}. 
The dashed lines show the limits of the instrumental range, also valid 
for figures a and c.
(c)~Experimental dynamic structure factor $S(Q,\omega)$ after correction for 
multiple scattering.
(d)~Dynamic many-body theory calculation of $S(Q,\omega)$.
Note that all the detailed features of the experimental data are reproduced. 
The units of the contour plots scale are meV$^{-1}.$ 
The intensity is cut off at 0.07\,meV$^{-1}$ in order to emphasize 
the multi-excitations region.
The apparent width of the Landau excitations in the experimental plot is due to 
the experimental resolution  
of 0.07\,meV, while the calculated Landau dispersion curve has been highlighted 
by a thick line.}
\label{Figure-1}
\end{figure}

\begin{figure}
\includegraphics[width=0.9\columnwidth]{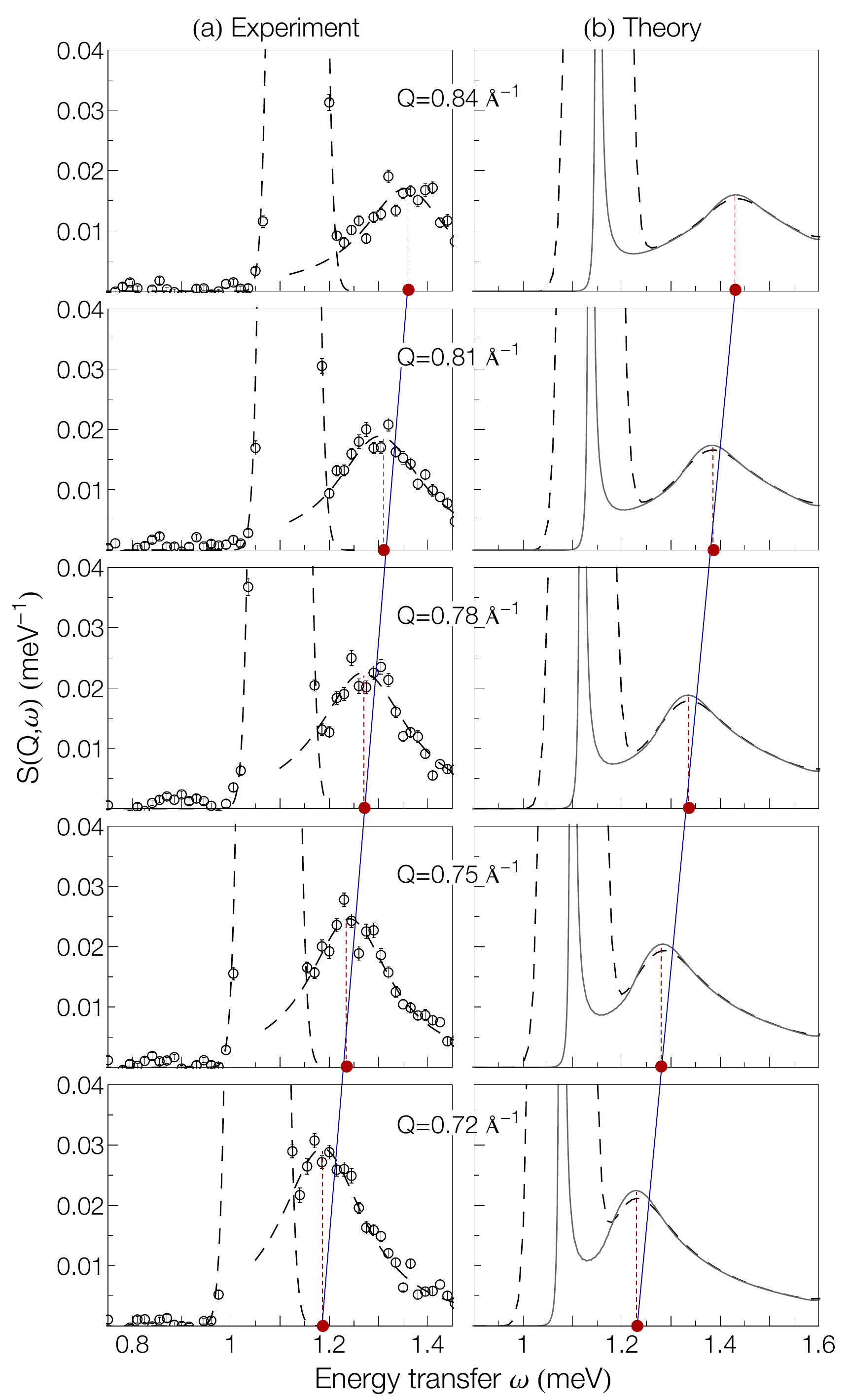}
\caption{(color online) (a)~Experimental dynamic structure factor $S(Q,\omega)$: 
spectra at different wave vectors Q in the phonon region. 
The dashed lines are Gaussian fits of the phonon peak (cut off) 
and the much smaller ``ghost phonon''.
The maxima of the ``ghost phonon'', indicated by vertical dashed lines, 
are located on the linear extension of the phonon 
dispersion (solid line).
(b)~Calculated spectra before and after convolution with the experimental 
resolution of 0.07\,meV, shown as solid and dashed lines, respectively. 
The experimental data are very well reproduced 
by the mode-mode coupling calculation. 
\label{Figure-2}}
\end{figure}

The quantity measured by a neutron spectrometer --the inelastic differential 
scattering cross section per target atom-- 
is proportional to the dynamic structure factor:  
\begin{equation}
\frac{\partial^{2}\sigma} {\partial\Omega~\partial\hbar\omega} = \frac{b^{2}_{c}} {\hbar} \frac{k'} {k} S(Q,\omega)  \nonumber 
\end{equation}
where $b_{c}$ is the bound atom coherent scattering length, $k$ and $k'$ the neutron 
wave vector before and after the scattering process, $Q$ the wave vector transfer and $\hbar\omega$ 
the energy transfer \cite{GlydeBook}. 
Standard data reduction routines \cite{lamp} were used to obtain the
dynamic structure factor from the neutron raw spectra.
The magnitude of $S(Q,\omega)$ was normalized by requiring that the single quasiparticle strength 
$Z(Q)\,=\,0.93$ for $Q\,=\,2.0$\,\AA$^{-1}$, a value  obtained from previous 
works \cite{CowleyWoods, GlydeBook,eomIII}.
Fig.\ \ref{Figure-1}a displays essentially the raw data, after the usual corrections.
The aluminum cell elastic background, measured before introducing the helium in the cell, 
was subtracted from the raw spectra.
This led to the noisy region seen in Fig.\ \ref{Figure-1}a near zero energy. 
We also subtracted the inelastic signal  
originating from scattering events involving 
the aluminum cell and the helium sample. Rotons, due 
to their high density of states, dominate these processes, and this contribution 
is only significant at the roton energy.
Since it is essentially Q-independent, it can be easily identified and removed.
The subtraction of this contribution spoils the accuracy of the data 
in a small range around the roton energy in regions of the spectrum where 
the signal is small. The effect can be seen 
if the intensity scale is considerably expanded, for instance as 
in Fig.\ \ref{Figure-2}.

While earlier neutron scattering experiments
\cite{CowleyWoods,Andersen92,Andersen94a,AndersenRoton} 
revealed the presence of
broad, rather featureless multi-particle excitation regions above the 
single-particle dispersion curve, 
the improved precision (and possibly the much lower temperature) in the present
experiment allowed us to observe a very rich structure in this region, 
with increasing weight at large wave
vectors, as seen in the measured $S(Q,\omega)$ shown 
in Fig.\ \ref{Figure-1}a. 

It is particularly important to distinguish the multi-particle 
excitations under investigation, 
which are an intrinsic property of helium, from multiple scattering.
The former arise when a neutron creates in a single process a high energy perturbation   
which can decay into two or more excitations, while 
the latter is a spurious effect, dependent on the sample size, where a single neutron 
creates two or more excitations in successive scattering events. 
Since the two kinds of processes fulfill 
the same kinematic conservation rules, and 
their contributions have similar intensity for typical sample sizes,  
subtracting multiple scattering from the raw data is essential when dealing with 
the multi-particle region of the spectrum. 

It is difficult in practice to determine this contribution experimentally \cite{Sears75}, 
for instance by using samples of 
different diameters, as one could naively believe. 
For this reason, we instead chose to perform an accurate calculation 
of the multiple scattering
contribution using simulation software   
\cite{Javier}, and we verified the results with two other simulation programs \cite{McStas,MScat}. 
All three Monte Carlo programs track successive neutrons in a
sample of given geometrical and physical characteristics,
in particular its scattering function $S(Q,\omega)$. They have been adapted to 
calculate multiple scattering in helium.
This contribution, shown in Fig.\ \ref{Figure-1}b, is found to be of the same order of magnitude 
as multi-particle excitations in all the region above 1.5\,meV. 
The ratio of the multiple scattering intensity 
over the total intensity is 1.8\,$\%$, a value in agreement with the calculation 
using the well-known semi-analytical method developed by Sears \cite{Sears75}. 
Multiple scattering was subtracted from the raw data, obtaining the corrected results for 
$S(Q,\omega)$ shown in Fig.\ \ref{Figure-1}c. The results shown in the figures clearly 
demonstrate that the new features we observe in this work are {\it not} caused 
by a multiple scattering artifact. 

Having obtained reliable data for the multi-particle region of $S(Q,\omega)$, we can 
claim the unambiguous experimental observation of three new features in $S(Q,\omega)$:
(i)~A feature that we have named {\it``ghost phonon''} 
for obvious reasons, since it appears 
as scattering strength extrapolating linearly from the phonon 
for wave vectors $0.6<Q<0.9$\,\AA$^{-1}$ ;
(ii)~A prominent branch-like feature that appears at about twice the maxon
energy; and (iii)~A sharp {\it threshold} at twice the roton energy that
extends to low wave vectors. 
Previous experimental works, including ours, show traces of the second feature,
at the limit of the experimental resolution. Comparing the plots of 
Ref. \cite{CowleyWoods,Andersen92,Andersen94a,AndersenRoton} 
with the results shown in Fig.\ \ref{Figure-1}c illustrates the magnitude 
of the improvement in neutron techniques over the last decade.
  
The kinematic possibility of the decay of a mode into two low-lying
quasiparticle excitations under energy and momentum conservation is a
necessary condition for these effects to occur, and the observed
features can be ascribed to modes decaying into pairs of excitations
of large spectral weight.  Even in the absence of a theory, it is
possible to combine pairs of single-particle excitations to obtain the
{\it position} of the main multi-excitation resonances in the dynamic
structure factor (2-Phonons, 2-Rotons, 2-Maxons and M+R resonances).
Obtaining the fine structure we observe, however, requires a
quantitative calculation of mode couplings.  Theoretical calculations
\cite{Saarela-TriesteBook,Saarela} based on early versions of the CBF
method gave a highly structured multi-excitation region in the dynamic
structure factor, which at first sight looks similar to the present
experimental results.  The similarity does not resist a more thorough
inspection, since even the single particle modes were only very
qualitatively reproduced.  In particular, the calculated
multi-excitations decay into Feynman modes instead of the true
single-particle excitations, leaving large gaps in the spectrum.
Also, the calculations predicted several additional features at high
energies that have not been found in experiments.  They provided
nevertheless an appealing example of the effects that could be
expected, and hence a motivation for further investigation of
multi-particle dynamics.

The challenge to the theory imposed by the present 
data was to obtain the magnitudes of the 
corresponding coupling matrix elements. 
They have been calculated in this work using 
Dynamic Many-body theory \cite{eomI,eomIII}, which can describe 
excitations with wavelengths 
comparable to the interparticle distance. This requires an appropriate 
treatment of correlations at atomic scales. 
In their pioneering work, Feynman and Cohen introduced pair 
fluctuations \cite{FeynmanBackflow}, which allowed them 
to take into account ``backflow'' effects. In order to calculate 
mode-mode couplings in a microscopic and quantitative manner, we include 
here $n-$particle fluctuations for all $n$.

The dynamic structure factor calculated for a number density of 
$\rho\,=\,0.022\,$\AA$^{-3}$ 
is  shown in Fig.\ \ref{Figure-1}d.
It is in quite satisfactory agreement with 
the experimental data. 
In addition to the well-known features of superfluid $^4$He, namely, 
the linear phonon dispersion relation, the maxon, and the
roton part of the spectrum turning eventually into the Pitaevskii
plateau \cite{Pitaevskii2Roton}, 
the calculated $S(Q,\omega)$ also shows the finer features 
revealed by our high-precision experiment (see Fig.\ \ref{Figure-1}c).

Our Dynamic Many-body calculations predict a ``ghost phonon'' which, 
as seen in Fig.\ \ref{Figure-1}, extends to about twice 
the wave vector up to which the dispersion relation is 
essentially linear. 
This can be shown explicitly by the calculation of the 
three-phonon interaction \cite{eomIII}. 
Fig.\ \ref{Figure-2} shows that the energy, strength, and 
shape of the calculated "ghost-phonon" agree very well 
with the experiment, even for wave vectors at atomic scale, 
of the order of 1\,\AA$^{-1}$. 

The mode above the maxon stems from interactions between
rotons and maxons. Modes with that frequency and a wave vector
close to that of the maxon decay predominantly into maxon and roton
excitations with momenta close to those of rotons and maxons, but
directed parallel and anti-parallel to the initial wave vector,
respectively. A discussion of the kinematic situation, estimated by
approximating the dispersion relation in the maxon and roton regions, 
is given in Ref. \cite{eomIII}.
This feature is clearly present both in the experiments and the theory, 
and Fig.\ \ref{Figure-1} c and d 
are very similar for energies as high as 2\,meV.

\begin{figure}
\includegraphics[width=1.0\columnwidth]{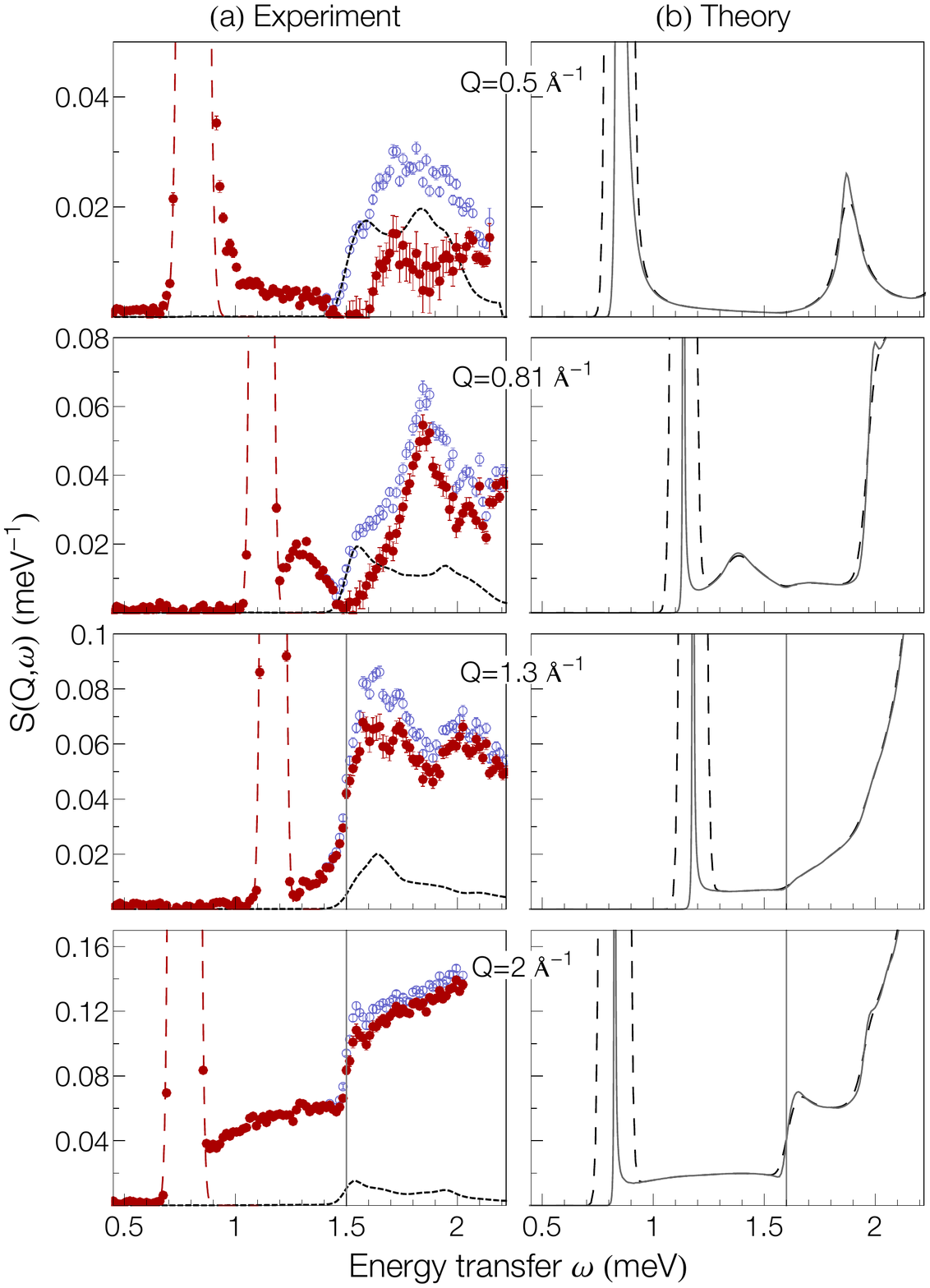}
\caption{(color online) (a) Experimental dynamic structure factor $S(Q,\omega)$: 
spectra for different wave vectors Q. 
Open circles: raw data; 
short-dashed line: calculated multiple scattering contribution; 
filled circles: $S(Q,\omega)$ corrected data (see text).
The dashed lines are Gaussian fits of the phonon-roton peaks (cut off).
(b) Theoretical dynamic structure factor spectra for the 
same wave vectors Q, before 
and after convolution with the experimental resolution of 0.07\,meV,  
shown as solid and dashed lines, respectively.
The plots have been chosen to emphasize differences between theory 
and experiment not readily observable in Fig.\,\ref{Figure-1}. 
Thin vertical lines: roton-roton threshold. 
\label{Figure-3}}
\end{figure}

A more quantitative comparison can be made on the spectra shown in 
Fig.\ \ref{Figure-3}.
The experimental spectrum at $Q\,=\,0.5$\,\AA$^{-1}$ is indeed very close  
to its theoretical analogue. At higher wave vectors, $Q\,=\,0.8$\,\AA$^{-1}$, however,
it is clear that only the low energy side of the experimental feature is 
reproduced by the theory. Contrarily to the situation encountered 
in the phonon region, 
where a very detailed calculation could be made, here the number 
of possible intermediate channels 
renders the calculation particularly demanding. 
The obtained semi-quantitative agreement is already quite gratifying. 

The decay of an excitation of energy
$2\Delta_{\text{r}}$ into two rotons with increasing angle
between them explains 
the plateau or threshold that we observe down to low wave vectors.
The effect, which has been extensively investigated \cite{Pitaevskii2Roton}, 
is discussed in detail in Ref. \cite{eomIII}.
The noteworthy observation is that 
this structure is observable, with comparable strength in
both experiments and theory (see Fig.\ \ref{Figure-3}), 
even below the maxon wave vector.

In conclusion, the dynamic structure factor of superfluid $^4$He, a canonical
model system for correlated boson physics, exhibits a rich
behavior at intermediate energies. We observe thresholds and fine 
structures delimited by kinematic constraints. The results 
are in remarkable agreement with the microscopic Dynamic
Many-body calculation. 
There are still some quantitative differences, observable above the roton energy for 
wave vectors smaller than the roton one, 
between 1 and 2\,\AA$^{-1}$. The experiments display a sizable response, which is 
present but weaker in the theory, as seen in 
Figs.\,\ref{Figure-1} and \ref{Figure-3}. 
The origin of the observed enhancement of this interesting mode remains  
an open question for theorists. 
There are also undeniable discrepancies with the theory at energies above 2 meV.
These can however be easily ascribed to higher order processes \cite{eomIII}; 
they are mostly 
structureless and do not lend themselves to a clear identification.
The combined experimental and theoretical work reported herein represents significant 
progress in the understanding of the effects of correlations on the dynamics 
of quantum fluids.

\begin{acknowledgments}
  We are grateful to X. Tonon for his help with the experiment, and 
	to E. Farhi for his help with the program McStas.
	This work was supported, in part, by the Austrian Science Fund FWF
  grant I602, the French grant ANR-2010-INTB-403-01 and the European
  Community Research Infrastructures under the FP7 Capacities Specific
  Programme, MICROKELVIN project number 228464.
\end{acknowledgments}

%


\begin{thebibliography}{29}%
\makeatletter
\providecommand \@ifxundefined [1]{%
 \@ifx{#1\undefined}
}%
\providecommand \@ifnum [1]{%
 \ifnum #1\expandafter \@firstoftwo
 \else \expandafter \@secondoftwo
 \fi
}%
\providecommand \@ifx [1]{%
 \ifx #1\expandafter \@firstoftwo
 \else \expandafter \@secondoftwo
 \fi
}%
\providecommand \natexlab [1]{#1}%
\providecommand \enquote  [1]{``#1''}%
\providecommand \bibnamefont  [1]{#1}%
\providecommand \bibfnamefont [1]{#1}%
\providecommand \citenamefont [1]{#1}%
\providecommand \href@noop [0]{\@secondoftwo}%
\providecommand \href [0]{\begingroup \@sanitize@url \@href}%
\providecommand \@href[1]{\@@startlink{#1}\@@href}%
\providecommand \@@href[1]{\endgroup#1\@@endlink}%
\providecommand \@sanitize@url [0]{\catcode `\\12\catcode `\$12\catcode
  `\&12\catcode `\#12\catcode `\^12\catcode `\_12\catcode `\%12\relax}%
\providecommand \@@startlink[1]{}%
\providecommand \@@endlink[0]{}%
\providecommand \url  [0]{\begingroup\@sanitize@url \@url }%
\providecommand \@url [1]{\endgroup\@href {#1}{\urlprefix }}%
\providecommand \urlprefix  [0]{URL }%
\providecommand \Eprint [0]{\href }%
\providecommand \doibase [0]{http://dx.doi.org/}%
\providecommand \selectlanguage [0]{\@gobble}%
\providecommand \bibinfo  [0]{\@secondoftwo}%
\providecommand \bibfield  [0]{\@secondoftwo}%
\providecommand \translation [1]{[#1]}%
\providecommand \BibitemOpen [0]{}%
\providecommand \bibitemStop [0]{}%
\providecommand \bibitemNoStop [0]{.\EOS\space}%
\providecommand \EOS [0]{\spacefactor3000\relax}%
\providecommand \BibitemShut  [1]{\csname bibitem#1\endcsname}%
\let\auto@bib@innerbib\@empty
\bibitem [{\citenamefont {Landau}(1941)}]{Landauroton}%
  \BibitemOpen
  \bibfield  {author} {\bibinfo {author} {\bibfnamefont {L.~D.}\ \bibnamefont
  {Landau}},\ }\href@noop {} {\bibfield  {journal} {\bibinfo  {journal} {J. of
  Phys. (Moscow)}\ }\textbf {\bibinfo {volume} {5}},\ \bibinfo {pages} {71}
  (\bibinfo {year} {1941})}\BibitemShut {NoStop}%
\bibitem [{\citenamefont {Landau}(1947)}]{Landauroton2}%
  \BibitemOpen
  \bibfield  {author} {\bibinfo {author} {\bibfnamefont {L.~D.}\ \bibnamefont
  {Landau}},\ }\href@noop {} {\bibfield  {journal} {\bibinfo  {journal} {J. of
  Phys. (Moscow)}\ }\textbf {\bibinfo {volume} {11}},\ \bibinfo {pages} {91}
  (\bibinfo {year} {1947})}\BibitemShut {NoStop}%
\bibitem [{\citenamefont {Feynman}\ and\ \citenamefont
  {Cohen}(1956)}]{FeynmanBackflow}%
  \BibitemOpen
  \bibfield  {author} {\bibinfo {author} {\bibfnamefont {R.~P.}\ \bibnamefont
  {Feynman}}\ and\ \bibinfo {author} {\bibfnamefont {M.}~\bibnamefont
  {Cohen}},\ }\href {\doibase 10.1103/PhysRev.102.1189} {\bibfield  {journal}
  {\bibinfo  {journal} {Phys. Rev.}\ }\textbf {\bibinfo {volume} {102}},\
  \bibinfo {pages} {1189} (\bibinfo {year} {1956})}\BibitemShut {NoStop}%
\bibitem [{\citenamefont {Macia}\ \emph {et~al.}(2012)\citenamefont {Macia},
  \citenamefont {Hufnagl}, \citenamefont {Mazzanti}, \citenamefont {Boronat},\
  and\ \citenamefont {Zillich}}]{PhysRevLett.109.235307}%
  \BibitemOpen
  \bibfield  {author} {\bibinfo {author} {\bibfnamefont {A.}~\bibnamefont
  {Macia}}, \bibinfo {author} {\bibfnamefont {D.}~\bibnamefont {Hufnagl}},
  \bibinfo {author} {\bibfnamefont {F.}~\bibnamefont {Mazzanti}}, \bibinfo
  {author} {\bibfnamefont {J.}~\bibnamefont {Boronat}}, \ and\ \bibinfo
  {author} {\bibfnamefont {R.~E.}\ \bibnamefont {Zillich}},\ }\href {\doibase
  10.1103/PhysRevLett.109.235307} {\bibfield  {journal} {\bibinfo  {journal}
  {Phys. Rev. Lett.}\ }\textbf {\bibinfo {volume} {109}},\ \bibinfo {pages}
  {235307} (\bibinfo {year} {2012})}\BibitemShut {NoStop}%
\bibitem [{\citenamefont {Bertaina}\ \emph {et~al.}(2016)\citenamefont
  {Bertaina}, \citenamefont {Motta}, \citenamefont {Rossi}, \citenamefont
  {Vitali},\ and\ \citenamefont {Galli}}]{1D-4He}%
  \BibitemOpen
  \bibfield  {author} {\bibinfo {author} {\bibfnamefont {G.}~\bibnamefont
  {Bertaina}}, \bibinfo {author} {\bibfnamefont {M.}~\bibnamefont {Motta}},
  \bibinfo {author} {\bibfnamefont {M.}~\bibnamefont {Rossi}}, \bibinfo
  {author} {\bibfnamefont {E.}~\bibnamefont {Vitali}}, \ and\ \bibinfo {author}
  {\bibfnamefont {D.~E.}\ \bibnamefont {Galli}},\ }\href {\doibase
  10.1103/PhysRevLett.116.135302} {\bibfield  {journal} {\bibinfo  {journal}
  {Phys. Rev. Lett.}\ }\textbf {\bibinfo {volume} {116}},\ \bibinfo {pages}
  {135302} (\bibinfo {year} {2016})}\BibitemShut {NoStop}%
\bibitem [{\citenamefont {Godfrin}\ \emph {et~al.}(2012)\citenamefont
  {Godfrin}, \citenamefont {Meschke}, \citenamefont {Lauter}, \citenamefont
  {Sultan}, \citenamefont {B{\"o}hm}, \citenamefont {Krotscheck},\ and\
  \citenamefont {Panholzer}}]{Nature2p2h}%
  \BibitemOpen
  \bibfield  {author} {\bibinfo {author} {\bibfnamefont {H.}~\bibnamefont
  {Godfrin}}, \bibinfo {author} {\bibfnamefont {M.}~\bibnamefont {Meschke}},
  \bibinfo {author} {\bibfnamefont {H.-J.}\ \bibnamefont {Lauter}}, \bibinfo
  {author} {\bibfnamefont {A.}~\bibnamefont {Sultan}}, \bibinfo {author}
  {\bibfnamefont {H.~M.}\ \bibnamefont {B{\"o}hm}}, \bibinfo {author}
  {\bibfnamefont {E.}~\bibnamefont {Krotscheck}}, \ and\ \bibinfo {author}
  {\bibfnamefont {M.}~\bibnamefont {Panholzer}},\ }\href {\doibase
  doi:10.1038/nature10919} {\bibfield  {journal} {\bibinfo  {journal} {Nature}\
  }\textbf {\bibinfo {volume} {483}},\ \bibinfo {pages} {576} (\bibinfo {year}
  {2012})}\BibitemShut {NoStop}%
\bibitem [{\citenamefont {Donnelly}\ \emph {et~al.}(1981)\citenamefont
  {Donnelly}, \citenamefont {Donnelly},\ and\ \citenamefont
  {Hills}}]{DonnellyDonnellyHills}%
  \BibitemOpen
  \bibfield  {author} {\bibinfo {author} {\bibfnamefont {R.~J.}\ \bibnamefont
  {Donnelly}}, \bibinfo {author} {\bibfnamefont {J.~A.}\ \bibnamefont
  {Donnelly}}, \ and\ \bibinfo {author} {\bibfnamefont {R.~N.}\ \bibnamefont
  {Hills}},\ }\href@noop {} {\bibfield  {journal} {\bibinfo  {journal} {J. Low
  Temp. Phys.}\ }\textbf {\bibinfo {volume} {44}},\ \bibinfo {pages} {471}
  (\bibinfo {year} {1981})}\BibitemShut {NoStop}%
\bibitem [{\citenamefont {Pines}\ and\ \citenamefont
  {Nozi{\`e}res}(1990)}]{PinesNoz}%
  \BibitemOpen
  \bibfield  {author} {\bibinfo {author} {\bibfnamefont {D.}~\bibnamefont
  {Pines}}\ and\ \bibinfo {author} {\bibfnamefont {P.}~\bibnamefont
  {Nozi{\`e}res}},\ }\href@noop {} {\emph {\bibinfo {title} {The theory of
  quantum liquids}}},\ Vol.~\bibinfo {volume} {I}\ (\bibinfo  {publisher}
  {Addison-Wesley},\ \bibinfo {year} {1990})\BibitemShut {NoStop}%
\bibitem [{\citenamefont {Glyde}(1994)}]{GlydeBook}%
  \BibitemOpen
  \bibfield  {author} {\bibinfo {author} {\bibfnamefont {H.~R.}\ \bibnamefont
  {Glyde}},\ }\href@noop {} {\emph {\bibinfo {title} {Excitations in liquid and
  solid helium}}}\ (\bibinfo  {publisher} {Clarendon Press Oxford},\ \bibinfo
  {year} {1994})\BibitemShut {NoStop}%
\bibitem [{\citenamefont {Cowley}\ and\ \citenamefont
  {Woods}(1971)}]{CowleyWoods}%
  \BibitemOpen
  \bibfield  {author} {\bibinfo {author} {\bibfnamefont {R.~A.}\ \bibnamefont
  {Cowley}}\ and\ \bibinfo {author} {\bibfnamefont {A.~D.~B.}\ \bibnamefont
  {Woods}},\ }\href {\doibase 10.1139/p71-021} {\bibfield  {journal} {\bibinfo
  {journal} {Can. J. Phys.}\ }\textbf {\bibinfo {volume} {49}},\ \bibinfo
  {pages} {177} (\bibinfo {year} {1971})},\ \Eprint
  {http://arxiv.org/abs/http://dx.doi.org/10.1139/p71-021}
  {http://dx.doi.org/10.1139/p71-021} \BibitemShut {NoStop}%
\bibitem [{\citenamefont {Andersen}\ \emph {et~al.}(1992)\citenamefont
  {Andersen}, \citenamefont {Stirling}, \citenamefont {Scherm}, \citenamefont
  {Stunault}, \citenamefont {F{\aa}k}, \citenamefont {Godfrin},\ and\
  \citenamefont {Dianoux}}]{Andersen92}%
  \BibitemOpen
  \bibfield  {author} {\bibinfo {author} {\bibfnamefont {K.~H.}\ \bibnamefont
  {Andersen}}, \bibinfo {author} {\bibfnamefont {W.~G.}\ \bibnamefont
  {Stirling}}, \bibinfo {author} {\bibfnamefont {R.}~\bibnamefont {Scherm}},
  \bibinfo {author} {\bibfnamefont {A.}~\bibnamefont {Stunault}}, \bibinfo
  {author} {\bibfnamefont {B.}~\bibnamefont {F{\aa}k}}, \bibinfo {author}
  {\bibfnamefont {H.}~\bibnamefont {Godfrin}}, \ and\ \bibinfo {author}
  {\bibfnamefont {A.}~\bibnamefont {Dianoux}},\ }\href@noop {} {\bibfield
  {journal} {\bibinfo  {journal} {Physica B: Condensed Matter}\ }\textbf
  {\bibinfo {volume} {180}},\ \bibinfo {pages} {851} (\bibinfo {year}
  {1992})}\BibitemShut {NoStop}%
\bibitem [{\citenamefont {Andersen}\ \emph {et~al.}(1994)\citenamefont
  {Andersen}, \citenamefont {Stirling}, \citenamefont {Scherm}, \citenamefont
  {Stunault}, \citenamefont {F{\aa}k}, \citenamefont {Godfrin},\ and\
  \citenamefont {Dianoux}}]{Andersen94a}%
  \BibitemOpen
  \bibfield  {author} {\bibinfo {author} {\bibfnamefont {K.~H.}\ \bibnamefont
  {Andersen}}, \bibinfo {author} {\bibfnamefont {W.~G.}\ \bibnamefont
  {Stirling}}, \bibinfo {author} {\bibfnamefont {R.}~\bibnamefont {Scherm}},
  \bibinfo {author} {\bibfnamefont {A.}~\bibnamefont {Stunault}}, \bibinfo
  {author} {\bibfnamefont {B.}~\bibnamefont {F{\aa}k}}, \bibinfo {author}
  {\bibfnamefont {H.}~\bibnamefont {Godfrin}}, \ and\ \bibinfo {author}
  {\bibfnamefont {A.~J.}\ \bibnamefont {Dianoux}},\ }\href@noop {} {\bibfield
  {journal} {\bibinfo  {journal} {J. Phys. Condens. Matter}\ }\textbf {\bibinfo
  {volume} {6}},\ \bibinfo {pages} {821} (\bibinfo {year} {1994})}\BibitemShut
  {NoStop}%
\bibitem [{\citenamefont {Gibbs}\ \emph {et~al.}(1999)\citenamefont {Gibbs},
  \citenamefont {Andersen}, \citenamefont {Stirling},\ and\ \citenamefont
  {Schober}}]{AndersenRoton}%
  \BibitemOpen
  \bibfield  {author} {\bibinfo {author} {\bibfnamefont {M.~R.}\ \bibnamefont
  {Gibbs}}, \bibinfo {author} {\bibfnamefont {K.~H.}\ \bibnamefont {Andersen}},
  \bibinfo {author} {\bibfnamefont {W.~G.}\ \bibnamefont {Stirling}}, \ and\
  \bibinfo {author} {\bibfnamefont {H.}~\bibnamefont {Schober}},\ }\href@noop
  {} {\bibfield  {journal} {\bibinfo  {journal} {J. Phys. Condens. Matter}\
  }\textbf {\bibinfo {volume} {11}},\ \bibinfo {pages} {603} (\bibinfo {year}
  {1999})}\BibitemShut {NoStop}%
\bibitem [{\citenamefont {Mezei}(1980)}]{Mezei80}%
  \BibitemOpen
  \bibfield  {author} {\bibinfo {author} {\bibfnamefont {F.}~\bibnamefont
  {Mezei}},\ }\href {\doibase 10.1103/PhysRevLett.44.1601} {\bibfield
  {journal} {\bibinfo  {journal} {Phys. Rev. Lett.}\ }\textbf {\bibinfo
  {volume} {44}},\ \bibinfo {pages} {1601} (\bibinfo {year}
  {1980})}\BibitemShut {NoStop}%
\bibitem [{\citenamefont {F\aa{}k}\ \emph {et~al.}(2012)\citenamefont
  {F\aa{}k}, \citenamefont {Keller}, \citenamefont {Zhitomirsky},\ and\
  \citenamefont {Chernyshev}}]{Fak12}%
  \BibitemOpen
  \bibfield  {author} {\bibinfo {author} {\bibfnamefont {B.}~\bibnamefont
  {F\aa{}k}}, \bibinfo {author} {\bibfnamefont {T.}~\bibnamefont {Keller}},
  \bibinfo {author} {\bibfnamefont {M.~E.}\ \bibnamefont {Zhitomirsky}}, \ and\
  \bibinfo {author} {\bibfnamefont {A.~L.}\ \bibnamefont {Chernyshev}},\ }\href
  {\doibase 10.1103/PhysRevLett.109.155305} {\bibfield  {journal} {\bibinfo
  {journal} {Phys. Rev. Lett.}\ }\textbf {\bibinfo {volume} {109}},\ \bibinfo
  {pages} {155305} (\bibinfo {year} {2012})}\BibitemShut {NoStop}%
\bibitem [{\citenamefont {Fabrocini}\ \emph {et~al.}(2002)\citenamefont
  {Fabrocini}, \citenamefont {Fantoni},\ and\ \citenamefont
  {Krotscheck}}]{TriesteBook}%
  \BibitemOpen
  \bibinfo {editor} {\bibfnamefont {A.}~\bibnamefont {Fabrocini}}, \bibinfo
  {editor} {\bibfnamefont {S.}~\bibnamefont {Fantoni}}, \ and\ \bibinfo
  {editor} {\bibfnamefont {E.}~\bibnamefont {Krotscheck}},\ eds.,\ \href@noop
  {} {\emph {\bibinfo {title} {Introduction to Modern Methods of Quantum
  Many--Body Theory and their Applications}}},\ \bibinfo {series} {Advances in
  Quantum Many--Body Theory}, Vol.~\bibinfo {volume} {7}\ (\bibinfo
  {publisher} {World Scientific},\ \bibinfo {address} {Singapore},\ \bibinfo
  {year} {2002})\BibitemShut {NoStop}%
\bibitem [{\citenamefont {Moroni}\ \emph {et~al.}(1998)\citenamefont {Moroni},
  \citenamefont {Galli}, \citenamefont {Fantoni},\ and\ \citenamefont
  {Reatto}}]{Moroni}%
  \BibitemOpen
  \bibfield  {author} {\bibinfo {author} {\bibfnamefont {S.}~\bibnamefont
  {Moroni}}, \bibinfo {author} {\bibfnamefont {D.~E.}\ \bibnamefont {Galli}},
  \bibinfo {author} {\bibfnamefont {S.}~\bibnamefont {Fantoni}}, \ and\
  \bibinfo {author} {\bibfnamefont {L.}~\bibnamefont {Reatto}},\ }\href
  {\doibase 10.1103/PhysRevB.58.909} {\bibfield  {journal} {\bibinfo  {journal}
  {Phys. Rev. B}\ }\textbf {\bibinfo {volume} {58}},\ \bibinfo {pages} {909}
  (\bibinfo {year} {1998})}\BibitemShut {NoStop}%
\bibitem [{\citenamefont {Vitali}\ \emph {et~al.}(2010)\citenamefont {Vitali},
  \citenamefont {Rossi}, \citenamefont {Reatto},\ and\ \citenamefont
  {Galli}}]{Vitali}%
  \BibitemOpen
  \bibfield  {author} {\bibinfo {author} {\bibfnamefont {E.}~\bibnamefont
  {Vitali}}, \bibinfo {author} {\bibfnamefont {M.}~\bibnamefont {Rossi}},
  \bibinfo {author} {\bibfnamefont {L.}~\bibnamefont {Reatto}}, \ and\ \bibinfo
  {author} {\bibfnamefont {D.~E.}\ \bibnamefont {Galli}},\ }\href {\doibase
  10.1103/PhysRevB.82.174510} {\bibfield  {journal} {\bibinfo  {journal} {Phys.
  Rev. B}\ }\textbf {\bibinfo {volume} {82}},\ \bibinfo {pages} {174510}
  (\bibinfo {year} {2010})}\BibitemShut {NoStop}%
\bibitem [{\citenamefont {Arrigoni}\ \emph {et~al.}(2013)\citenamefont
  {Arrigoni}, \citenamefont {Vitali}, \citenamefont {Galli},\ and\
  \citenamefont {Reatto}}]{Arrigoni}%
  \BibitemOpen
  \bibfield  {author} {\bibinfo {author} {\bibfnamefont {F.}~\bibnamefont
  {Arrigoni}}, \bibinfo {author} {\bibfnamefont {E.}~\bibnamefont {Vitali}},
  \bibinfo {author} {\bibfnamefont {D.~E.}\ \bibnamefont {Galli}}, \ and\
  \bibinfo {author} {\bibfnamefont {L.}~\bibnamefont {Reatto}},\ }\href
  {\doibase http://dx.doi.org/10.1063/1.4821079} {\bibfield  {journal}
  {\bibinfo  {journal} {Low Temperature Physics}\ }\textbf {\bibinfo {volume}
  {39}},\ \bibinfo {pages} {793} (\bibinfo {year} {2013})},\ \bibinfo {note}
  {[Fiz. Nizk. Temp. 39, 1021 (2013)]}\BibitemShut {NoStop}%
\bibitem [{\citenamefont {Campbell}\ \emph {et~al.}(2015)\citenamefont
  {Campbell}, \citenamefont {Krotscheck},\ and\ \citenamefont
  {Lichtenegger}}]{eomIII}%
  \BibitemOpen
  \bibfield  {author} {\bibinfo {author} {\bibfnamefont {C.~E.}\ \bibnamefont
  {Campbell}}, \bibinfo {author} {\bibfnamefont {E.}~\bibnamefont
  {Krotscheck}}, \ and\ \bibinfo {author} {\bibfnamefont {T.}~\bibnamefont
  {Lichtenegger}},\ }\href {\doibase 10.1103/PhysRevB.91.184510} {\bibfield
  {journal} {\bibinfo  {journal} {Phys. Rev. B}\ }\textbf {\bibinfo {volume}
  {91}},\ \bibinfo {pages} {184510} (\bibinfo {year} {2015})}\BibitemShut
  {NoStop}%
\bibitem [{lam()}]{lamp}%
  \BibitemOpen
  \href {http://www.ill.eu/data\_treat/lamp} {\ }\bibinfo {note} {{\sf LAMP}
  (Large Array Manipulation Program) {
  http://www.ill.eu/data\_treat/lamp}}\BibitemShut {NoStop}%
\bibitem [{\citenamefont {Sears}(1975)}]{Sears75}%
  \BibitemOpen
  \bibfield  {author} {\bibinfo {author} {\bibfnamefont {V.~F.}\ \bibnamefont
  {Sears}},\ }\href@noop {} {\bibfield  {journal} {\bibinfo  {journal} {Adv.
  Phys.}\ }\textbf {\bibinfo {volume} {24}},\ \bibinfo {pages} {1} (\bibinfo
  {year} {1975})}\BibitemShut {NoStop}%
\bibitem [{\citenamefont {Dawidowski}\ \emph {et~al.}(1998)\citenamefont
  {Dawidowski}, \citenamefont {Bermejo},\ and\ \citenamefont
  {Granada}}]{Javier}%
  \BibitemOpen
  \bibfield  {author} {\bibinfo {author} {\bibfnamefont {J.}~\bibnamefont
  {Dawidowski}}, \bibinfo {author} {\bibfnamefont {F.~J.}\ \bibnamefont
  {Bermejo}}, \ and\ \bibinfo {author} {\bibfnamefont {J.~R.}\ \bibnamefont
  {Granada}},\ }\href {\doibase 10.1103/PhysRevB.58.706} {\bibfield  {journal}
  {\bibinfo  {journal} {Phys. Rev. B}\ }\textbf {\bibinfo {volume} {58}},\
  \bibinfo {pages} {706} (\bibinfo {year} {1998})}\BibitemShut {NoStop}%
\bibitem [{\citenamefont {Lefmann}\ and\ \citenamefont
  {Nielsen}(1999)}]{McStas}%
  \BibitemOpen
  \bibfield  {author} {\bibinfo {author} {\bibfnamefont {K.}~\bibnamefont
  {Lefmann}}\ and\ \bibinfo {author} {\bibfnamefont {K.}~\bibnamefont
  {Nielsen}},\ }\href {http://www.mcstas.org} {\bibfield  {journal} {\bibinfo
  {journal} {Neutron news}\ }\textbf {\bibinfo {volume} {10}},\ \bibinfo
  {pages} {20} (\bibinfo {year} {1999})}\BibitemShut {NoStop}%
\bibitem [{\citenamefont {Copley}\ \emph {et~al.}(1986)\citenamefont {Copley},
  \citenamefont {Verkerk}, \citenamefont {van Well},\ and\ \citenamefont
  {Fredrikze}}]{MScat}%
  \BibitemOpen
  \bibfield  {author} {\bibinfo {author} {\bibfnamefont {J.~R.~D.}\
  \bibnamefont {Copley}}, \bibinfo {author} {\bibfnamefont {P.}~\bibnamefont
  {Verkerk}}, \bibinfo {author} {\bibfnamefont {A.~A.}\ \bibnamefont {van
  Well}}, \ and\ \bibinfo {author} {\bibfnamefont {H.}~\bibnamefont
  {Fredrikze}},\ }\href@noop {} {\bibfield  {journal} {\bibinfo  {journal}
  {Comput. phys. commun.}\ }\textbf {\bibinfo {volume} {40}},\ \bibinfo {pages}
  {337} (\bibinfo {year} {1986})}\BibitemShut {NoStop}%
\bibitem [{\citenamefont {Saarela}(2002)}]{Saarela-TriesteBook}%
  \BibitemOpen
  \bibfield  {author} {\bibinfo {author} {\bibfnamefont {M.}~\bibnamefont
  {Saarela}},\ }\href@noop {} {}\ (\bibinfo {year} {2002})\ pp.\ \bibinfo
  {pages} {205--264},\ \bibinfo {note} {in Ref.
  {\cite{TriesteBook}}}\BibitemShut {NoStop}%
\bibitem [{\citenamefont {Saarela}(1986)}]{Saarela}%
  \BibitemOpen
  \bibfield  {author} {\bibinfo {author} {\bibfnamefont {M.}~\bibnamefont
  {Saarela}},\ }\href@noop {} {\bibfield  {journal} {\bibinfo  {journal} {Phys.
  Rev. B}\ }\textbf {\bibinfo {volume} {33}},\ \bibinfo {pages} {4596}
  (\bibinfo {year} {1986})}\BibitemShut {NoStop}%
\bibitem [{\citenamefont {Campbell}\ and\ \citenamefont
  {Krotscheck}(2009)}]{eomI}%
  \BibitemOpen
  \bibfield  {author} {\bibinfo {author} {\bibfnamefont {C.~E.}\ \bibnamefont
  {Campbell}}\ and\ \bibinfo {author} {\bibfnamefont {E.}~\bibnamefont
  {Krotscheck}},\ }\href {\doibase 10.1103/PhysRevB.80.174501} {\bibfield
  {journal} {\bibinfo  {journal} {Phys. Rev. B}\ }\textbf {\bibinfo {volume}
  {80}},\ \bibinfo {pages} {174501} (\bibinfo {year} {2009})}\BibitemShut
  {NoStop}%
\bibitem [{\citenamefont {Pitaevskii}(1959)}]{Pitaevskii2Roton}%
  \BibitemOpen
  \bibfield  {author} {\bibinfo {author} {\bibfnamefont {L.~P.}\ \bibnamefont
  {Pitaevskii}},\ }\href@noop {} {\bibfield  {journal} {\bibinfo  {journal}
  {Zh. Eksp. Theor. Fiz.}\ }\textbf {\bibinfo {volume} {36}},\ \bibinfo {pages}
  {1168} (\bibinfo {year} {1959})},\ \bibinfo {note} {[Sov. Phys. JETP {\bf 9},
  830 (1959)]}\BibitemShut {NoStop}%
\end{thebibliography}

\end{document}